\documentclass[acmlarge, sigconf]{acmart}

\AtBeginDocument{%
  }

\setcopyright{acmcopyright}
\copyrightyear{2023}
\acmYear{2023}
\acmDOI{XXXXXXX.XXXXXXX}

\usepackage{enumitem}
\usepackage[normalem]{ulem}
\usepackage{multirow}
\usepackage{siunitx}
\usepackage{adjustbox}

\begin{document}

\title{On the Effectiveness of Creating Conversational Agent Personalities Through Prompting}

\author{Heng Gu}
\affiliation{%
  \institution{Delft University of Technology}
  \city{Delft}
  \country{Netherlands}}
\email{h.gu@tudelft.nl}

\author{Chadha Degachi}
\affiliation{%
  \institution{Delft University of Technology}
  \city{Delft}
  \country{Netherlands}}
\email{c.degachi@tudelft.nl}

\author{Uğur Genç}
\affiliation{%
  \institution{Delft University of Technology}
  \city{Delft}
  \country{Netherlands}}
\email{u.genc@tudelft.nl}

\author{Senthil	Chandrasegaran}
\affiliation{%
  \institution{Delft University of Technology}
  \city{Delft}
  \country{Netherlands}}
\email{r.s.k.chandrasegaran@tudelft.nl}

\author{Himanshu Verma}
\affiliation{%
  \institution{Delft University of Technology}
  \city{Delft}
  \country{Netherlands}}
\email{h.verma@tudelft.nl}

\begin{abstract}
 In this work, we report on the effectiveness of our efforts to tailor the personality and conversational style of a conversational agent based on GPT-3.5 and GPT-4 through prompts. We use three personality dimensions with two levels each to create eight conversational agents archetypes. Ten conversations were collected per chatbot, of ten exchanges each, generating 1600 exchanges across GPT-3.5 and GPT-4. Using Linguistic Inquiry and Word Count (LIWC) analysis, we compared the eight agents on language elements including clout, authenticity, and emotion. Four language cues were significantly distinguishing in GPT-3.5, while twelve were distinguishing in GPT-4. With thirteen out of a total nineteen cues in LIWC appearing as significantly distinguishing, our results suggest possible novel prompting approaches may be needed to better suit the creation and evaluation of persistent conversational agent personalities or language styles. 
\end{abstract}

\ccsdesc[500]{Computing methodologies~Natural language generation}

\keywords{LLMs, Conversational Agents, Sentiment Analysis, Personality}

\maketitle

\section{Introduction}
Tailoring conversational agents (CAs) to express personality has long been a subject of interest for researchers. The perceived friendliness and (in)formality of a CA affect the human factors of user-chatbot interaction, including dimensions such as trust~\cite{folstad2018chatbottrust, xiao2020personalityaligment}, engagement~\cite{fadhil2018emojis}, and acceptance~\cite{sanny2020acceptance}. CAs with personality create a more consistent user experience, and can even improve the overall user experience~\cite{thies2017personality, smestad2018botpersonality}. Some researchers have even worked to create complementary alignments of CA-user personalities for improved user experience ~\cite{LiewTan2016personalitymatch, ISBISTER2000personalitymatch}.

The link between personality and language style is well established. For example, introverted speakers tend more towards formal language, with fewer exaggerations and more hedging, or tentative, phrasing~\cite{furnham1999extra, Heylighen2002extra, Oberlander2004extra, Pennebaker1999extra}. In prior work, such language markers have been used in a rule-based system to generate text expressing associated personalities to some success~\cite{mairesse2010personality}. Language use has also been linked to outcomes of negotiation and persuasion: for instance, language style in solicitations for charitable giving has been shown to be a predictor of the donation or assistance given~\cite{mitra2014language}. It thus follows that personality expressed through language in CAs can influence the outcome of an end-users' discussion with the CA. 

Recent advances in text generation using Large Language Models (LLMs), present a unique opportunity for non-rule-based, heuristic approaches to express personality through language. Yet, within this field, a conspicuous gap emerges when it comes to the capacity of LLMs to generate language styles corresponding to personality archetypes. Though the use of LLMs to power CAs is on the rise~\cite{basae2023llmbot, ross2023llmbot}, only one study~\cite{jiang2023personallm} begins to investigate the possibilities of personality generation in GPT-3.5, but little work quantifies LLMs' fidelity to, or linguistic variances influenced by, personality prompts over the length of a conversation. 

There have been studies in persuasive technologies and language design, in fields such as public health communication and marketing~\cite{Bai2023persuasion, Karinshak2023persuasion}.
However, these predominantly concentrate on direct user feedback or discernible shifts in persuasion outcomes to evaluate the effectiveness of linguistic element tailoring, rather than on
quantifying the actual variance and consistency of their chosen language cues.

In this work, we report on an approach to express personality through language in a goal-oriented CA. Specifically, we investigate the expression of personality communicated through language in the context of solicitation for charitable giving. We leverage prior research on charitable giving~\cite{mitra2014language} showing that the potential donor's engagement with and response to the associated cause is sensitive to the language used in solicitation. We use literature on charitable giving~\cite{small2007sympathy, chang2010effects, fatkhiyati2019rhetorical, chen2019multi} to derive three dimensions of personality that could be inferred from the language of solicitation. These are (1) \textsc{Attitude}, i.e., optimistic vs pessimistic, (2) \textsc{Authority}, i.e., authoritative vs submissive, and (3) \textsc{Reasoning}, i.e., analytical vs affective. 

Using ``prompting'', a technique that is used to customize the output of LLMs using natural-language guidelines~\cite{white2023prompt}, we express the above personalities, first in a CA powered by the Generative Pre-Trained Transformer (GPT)-3.5 LLM, and then by the GPT-4 LLM. Finally, we use Linguistic Inquiry and Word Count (LIWC)~\cite{boyd2022development}---a tool that uses psycholinguistic dictionary categories associating the use of certain words with certain psychological processes---to answer the following research question:

\begin{itemize}[leftmargin=2em]
    \item[\textbf{RQ.}] Does language generated by LLMs---using prompt engineering to express certain personalities---in conversational situations align with existing associations between said personalities and language use?
\end{itemize}

Our results show that GPT-4 outperforms GPT-3.5 in its ability to adhere to prompted personality language characteristics.
Using nineteen relevant dictionary categories in LIWC as potentially distinguishing characteristics of these personalities, we find that text from CAs powered by the GPT-4 LLM shows significant distinction between personalities along twelve characteristics, while GPT-3.5 shows the same for only four characteristics. However, given that we investigated a total of nineteen LIWC categories, we still see considerable room for improvement in the performance of the both models. Thus, we also provide recommendations for possible novel prompting strategies better suited to persistent personality generation. Moreover, through this study, we present a novel approach to CA personality design which diverges from the psychological models used in past work, to focus on real-world domain knowledge and tasks for the extraction of relevant personality traits.

\section{Related Work}
\subsection{Language and Personality}
A few theories of psychology and human behaviour connect language use and personality.
The concept of ``words as attention'' is one wherein researchers argue the language choice in people is deliberate and therefore reflective, to some extent, of their focus~\cite{Stone1966geninquire, Boyd2021langanalysis}.
The work of ~\citet{Wegner1987words} validates this framing to an extent, showing that when prompted to output a verbal stream of consciousness while avoiding thinking of a ``white bear'', participants often struggled and failed.
Other theories posit that language use indicates how individuals process information~\cite{Tausczik2010words}, or the intensity of their attitude towards a given subject~\cite{Stone1966geninquire}.
Originally, this style of analysis was carried out through manual, qualitative coding of text.

Computational methods of predicting personality from text and language cues also exist, with dictionary-based approaches such as LIWC among the most popular approaches.
In their work, ~\citet{Koutsoumpis2022liwcpersonality} show that twenty categories of language extracted from text by LIWC can be associated with the dimensions of the Big Five personality model in both self- and observer-reported personality traits.
For example, emotional stability was associated with the use of first-person singular pronouns, and conscientiousness with swear words, anger, and negative emotions.
Similarly, ~\citet{zhou2021sitempathy} and ~\citet{Lord2015sitempathy} use LIWC to analyse the situational empathy invoked in participants by distressing situations from text responses. Situational empathy can give us insight into user dispositional, or trait empathy, a stable element of personality, which has strong influence on both situational empathy and charitable giving~\cite{zhou2021sitempathy}.


\subsection{Language and Persuasion}
The effectiveness of persuasive messaging is known to relate to its relevance and personal appeal to the intended target~\cite{Petty1986elm}. Part of this personalisation is achieved through language. ~\citet{Pfeiffer2023charity} find that language formality in donation solicitation is positively correlated with charitable giving, for example. ~\citet{Ye2022enviromental} found figurative, as opposed to literal language, improved user's ability to visualize in the context of plant-based meat consumption and in turn increased favourable consumer behaviour. Lastly, ~\citet{Bai2023persuasion} show that statements incorporating rhetorical devices such as antanagoge, anaphora, and rhetorical questions were most influential in predicting text persuasiveness using machine learning models. 

In their work ~\citet{wang2020persuasion} developed a CA for persuasive messaging in the domain of charitable giving. The researchers used deep learning techniques to equip their agent with a variety of donation solicitation strategies expressed through language, including “rational” vs “emotional” appeals. The study concluded that the effect of a given donation strategy was moderated by the user's own personality, for example extroverted users were more likely to be persuaded by emotional donation appeals than others. Similar efforts using LLM technology do not yet exist.

\subsection{Personable Conversational Agents}
~\citet{mairesse2010personality}, and others since ~\cite{Harrison2019personalitynlg, Ruane2021personalityperception, Kuhail2022personalityadvisor} have studied personality-based natural language generation, as well as dialogue design, and the perception thereof. Much of this past work relies on the aforementioned Big-Five personality model, and creates associations between personality and language cues using linguistics and psychology literature. In a few cases, other psychological personality models, such as Myers-Briggs, and other language cue extraction methods, such as machine learning and sentiment analysis ~\cite{fernau2022interspeech} are employed. Few studies have worked to craft CA personalities in relation to the context and domain in which it is deployed. Moreover, these studies used a variety of approaches for implementing their conversational agents including deep learning and hand-crafted dialogue, however, few have investigated Large Language Models in this context. Given the performance improvement LLMs offer over previous NLG approaches~\cite{openai2023gpt4}, this is a great oversight.

As mentioned, one study~\cite{jiang2023personallm} so far has taken advantage of the rise of LLMs across the natural language generation space, to investigate whether these models could be prompted to create agents with personality. In their work, ~\citet{jiang2023personallm} used Big-Five personality dimensions, as well as gender, to prompt a GPT-3.5 agent to create an 800-word childhood story analysed using LIWC language elements. Though the authors concluded that the personas they created were significantly different in the LIWC language categories they exhibited, they did not investigate whether this pattern would hold in naturalistic longer human-agent interaction settings such as conversation. ~\citet{serapiogarcía2023personality} use a somewhat similar methodology to evaluate the PaLM models~\cite{chowdhery2022palm} in their ability to embody personality, using prompt engineering to create personas and asking the CA to answer psychometric personality tests. They similarly concluded CAs to be cable of consistent personality embodiment. 

\section{Method}
We prompt-engineered a series of charity solicitation CAs employing the GPT-3.5 and GPT-4 models.
The two models with similar architectures but different sizes and parameters were chosen for comparison with prior studies focusing primarily on aptitudes across various human expertise categories, such as through standardized exams \cite{bommarito2022gpt, liu2023evaluating, ateia2023chatgpt, koubaa2023gpt}, 
we believe it novel to compare the CA outputs through the lens of inguistic outputs.
These CAs, representing a fabricated charity organization named the ``Wildlife Bridge Foundation,'' were designed to simulate a solicitation event with potential donors. 

Building on prior research on effective charity solicitation, we integrated variations in the CA personality and solicitation strategy across three dimensions: \textit{Attitude}, i.e., optimistic vs pessimistic~\cite{small2007sympathy}, \textit{Authority}, i.e., authoritative vs submissive~\cite{chang2010effects, chen2019multi}, and \textit{Reasoning}, i.e., analytical vs affective~\cite{fatkhiyati2019rhetorical}, reflected in the language used to persuade the potential donor.
These three dimensions---each with two polar opposite attributes---combinatirially resulted in  eight ($2^3$) distinct CA personalities.
We powered one set of all eight CA personalities with the GPT-3.5 model and another set with the GPT-4 model. 

We collected the ten most, and least, popular petitions from Avaaz and Change.org, two popular online petition platforms, to analyse with LIWC, identifying the most frequently used language categories and psychological processes within. These characteristics included \textit{clout}, \textit{tone}, \textit{authenticity}, \textit{emotion}, \textit{cognition}, and others. Once identified, we grouped these characteristics along the three personality scales derived from  literature.
For example, authority was associated with the LIWC categories of clout and authenticity, and an authoritative text may have used more words and phrases such as “must”, “have to”, and “should”, in contrast to a submissive text which used words and phrases such as “if it's okay with you”, “maybe”, and “if you don't mind”.
Details of each personality dimension and associated LIWC categories are described in Sec.~\ref{sec:measures}.

\subsection{Prompt Engineering}
We first designed a core prompt with modifiable slots for varying solicitor traits. Grounded in advanced prompt design methodologies \cite{white2023prompt}, this foundational prompt encompasses four principal elements:
\begin{enumerate}
    \item \textbf{Task:} Act as a charity solicitor for… 
    \item \textbf{Goal:} Get speaker to donate… 
    \item \textbf{Rules:} Do not provide URLS, keep response short... 
    \item \textbf{Persona:} The solicitor's name is [NAME], personality: [optimistic/pessimistic], and [Authoritativeness/Submissiveness]. Only speak as Alex from now on. Use [Emotion/Logic-based reasoning] to convince the donor.
\end{enumerate}

\subsection{Benchmark Test}
To assess the performance of the 16 uniquely personalized CAs (8 GPT-3.5 and 8 GPT-4), we devised a standardized script composed of the same 10 dialogue excerpts,
which were queried to each CA over 10 sessions, resulting in 160 unique conversations and 1,600 pieces of generated responses, 100 per CA. 

This set of interactions was designed based on internal pilot tests simulating potential donor conversations.
We distilled the most pertinent questions and responses to form this script.

\section{Analysis}
Following synthesis of this benchmarking sample dataset of generated responses, our primary objective was to evaluate their consistency relative to our prompts indicating the desired output qualities. 
We utilized LIWC to assess the cognitive and emotional attributes of the generated text, measured along dictionary categories that we deemed relevant to the personality dimensions.

\subsection{Measures}
\label{sec:measures}

LIWC's psycholinguistic dictionary categories~\cite{boyd2022development} provided us with the means to gauge the linguistic attributes of each agent's outputs.

Specifically, we focused on LIWC categories known to reflect our three chosen traits:
\begin{description}
    \item[Attitude:] 
    For this trait, which could take the attribute of either \textit{optimistic} or \textit{pessimistic}, we looked at linguistic elements pointing towards sentiment. 
    In the current version of LIWC, \textit{optimism} is a subcategory of Affect~\cite{mcdonnell2015comparison}. 
    This parallels the \textit{Tone} categories (\textit{tone\_pos}, \textit{tone\_neg}) which are indicative of sentiment of the text as opposed to embodied emotions \cite{boyd2022development}. 
    Positive emotion(\textit{Emo\_Pos}) and tentativeness(\textit{tentat}) as indicators have also been used for optimism \cite{gasper2020differentiating}. 
    Additionally, based on the psychological correlates outlined by LIWC guidelines, 
    we also monitored \textit{FutureTense} representing future and goal-oriented words, and markers of \textit{Anxiety} indicating future-oriented emotions (e.g., words like worried, fearful, nervous) ~\cite{tausczik2010psychological, boyd2022development}.
    
    \item[Authority:] 
    For this trait, the attributes of which could be \textit{authoritative} or \textit{submissive}, we looked at linguistic elements pointing towards status, dominance, and social hierarchy. 
    Clout serves as an aggregate category representing social status and power dynamics, with subcategories of personal pronouns (\textit{ppron}) indicative of these dynamics associated with authority~\cite{buljan2020large, lumontod2020seeing}.
    Additionally, according to the psychometric guidelines, \textit{certitude} and absolutist language (\textit{All-none}) could also be an indicator of authoritative language, and \textit{Assent} category could be an indicator of submissiveness~\cite{boyd2022development}. 
    
    \item[Reasoning: ]
    In the context of solicitation traits, \textit{analytical} specifically refers to a solicitation strategy of providing statistical evidence, and \textit{affective} refers to providing individualized details~\cite{small2013sympathy, fatkhiyati2019rhetorical}. 
    For this trait we looked at linguistic elements pointing to thinking styles and cognitive mechanism words (\textit{Cognition, Analytic, Quantity, Numbers}), as well as the presence of emotional words  (\textit{Emotion, Affect, Authentic})~\cite{tausczik2010psychological, boyd2022development}.
\end{description}

\subsection{Procedure}
Given the selected traits, interaction effects between them are anticipated. Our objective was to discern the parts-worth effect attributable to the prompted solicitor parameters of Attitude, Reasoning, and Authority. 
To accomplish this, we employed conjoint analysis, aiming to deconstruct the individual impact of these traits on the output linguistic quality. 
The conceived equation is a linear regression model which relates the LIWC category values to the factors of Attitude, Authority, and Reasoning prompt components, along with their interaction effects.


The equation representing this relationship is:
\begin{equation}
\begin{split}
M_{LIWC} = & \beta_0 + \beta_1(\text{Attitude})  + \beta_2(\text{Authority}) + \\
 & + \beta_3(\text{Reasoning}) + \beta_4(\text{Attitude} \times \text{Authority}) \\
 & + \beta_5(\text{Attitude} \times \text{Reasoning}) \\
 & + \beta_6(\text{Reasoning} \times \text{Authority}) \\
 & + \beta_7(\text{Attitude} \times\text{Authority} \times \text{Reasoning}) + \epsilon
\end{split}
\end{equation}

\begin{table*}[t]
\centering
\small 
\renewcommand{\arraystretch}{0.7} 
\caption{Effect of Prompt Factors on LIWC Categories}
\vspace{-1mm}
\label{tab:LIWC_measures}
\begin{adjustbox}{center}
\begin{tabular}{llrrrrrrrrrrrrrrrr}
\toprule
 &
\multirow{2}{*}{LIWC} & \multicolumn{8}{c}{GPT3.5} & \multicolumn{8}{c}{GPT4} \\
\cmidrule(lr){3-10}
\cmidrule(lr){11-18}
 &
Category &
$R^2$ &
\(\beta_1\) &
\(\beta_2\) &
\(\beta_3\) &
\(\beta_4\) &
\(\beta_5\) &
\(\beta_6\) &
\(\beta_7\) &
$R^2$ &
\(\beta_1\) &
\(\beta_2\) &
\(\beta_3\) &
\(\beta_4\) &
\(\beta_5\) &
\(\beta_6\) &
\(\beta_7\) \\

\midrule
\parbox[t]{2mm}{\multirow{6}{*}{\rotatebox[origin=c]{90}{Authority}}} &
 Clout &
 0.040 &
   &
 - &
   &
 - &
   &
 - &
 - &
 0.270 &
   &
 **1.800 &
   &
 *-1.250 &
   &
 *-1.129 &
 - \\
 
   &
Authentic &
 0.083 &
   &
 - &
   &
 - &
   &
 - &
 - &
 0.120 &
   &
 - &
   &
 - &
   &
 - &
 - \\
 
   &
Ppron &
 0.228 &
   &
 - &
   &
 - &
   &
 - &
 - &
 0.112 &
   &
 - &
   &
 - &
   &
 - &
 - \\
 
   &
Certitude &
 0.096 &
   &
 - &
   &
- &
   &
 - &
 - &
 0.168 &
   &
 *-0.200 &
   &
 - &
   &
 - &
 - \\
 
   &
Allnone &
 0.227 &
   &
 - &
   &
 - &
   &
 - &
 - &
 0.168 &
   &
 *-0.148 &
   &
 *-0.130&
   &
 - &
 - \\
 
   &
Assent &
 0.987 &
   &
 **0.211 &
   &
 **0.217 &
   &
 - &
 - &
 0.967 &
   &
 *0.424 &
   &
 **0.406 &
   &
 **-0.447 &
 **-0.346 \\

\midrule
\parbox[t]{2mm}{\multirow{6}{*}{\rotatebox[origin=c]{90}{Attitude}}} &
 Tone &
 0.315 & 
 *0.334 & 
   & 
   & 
 *0.363 & 
 **0.448 & 
   & 
 - & 
 
 0.283 &
 **1.278 &
   &
   &
 *0.792 &
 - &
   &
 - \\
 
 &
 Affect &
 0.123 &
 - &
   &
   &
 *0.193 &
 - &
   &
 - &
 
 0.273 &
 **0.688 &
   &
   &
 - &
 - &
   &
 - \\
 
 &
 Tone\_Pos &
 0.161 &
 - &
   &
   &
 *0.234 &
 - &
   &
 - &
 
 0.402 &
 **0.997 &
   &
   &
 *0.325 &
 - &
   &
 - \\
 
 &
 Tone\_Neg &
 0.281 &
 **-0.082 &
   &
   &
 - &
 - &
   &
 - &
 
 0.338 &
 *-0.308 &
   &
   &
 - &
 - &
   &
 - \\
 
 &
 Emo\_Anx &
 0.089 &
 - &
   &
   &
 - &
 - &
   &
 - &
 
 0.257 &
 **-0.047 &
   &
   &
 *-0.029 &
 - &
   &
 - \\

\midrule
\parbox[t]{2mm}{\multirow{6}{*}{\rotatebox[origin=c]{90}{Reasoning}}}  &
 Cognition &
 0.097 & 
   &
   &
    - &
   &
 - &
 - &
 - &
 
 0.336 &
   &
   &
 *0.527 &
   &
-  &
 - &
 - \\
 
 &
 Analytic &
 0.201 & 
   &
   &
**-1.750 &
   &
 - &
 - &
 - &
 
 0.142 &
   &
   &
 - &
   &
-  &
 - &
 - \\
 
 &
 Authentic &
 0.083 & 
   &
   &
    - &
   &
 - &
 - &
 - &
 
 0.119 &
   &
   &
 - &
   &
-  &
 - &
 - \\
 
 &
 Quantity &
 0.136 & 
   &
   &
    - &
   &
 - &
 *-0.147 &
 - &
 
 0.129 &
   &
   &
 *0.291 &
   &
-  &
 - &
 - \\
 
 &
 Numbers &
 0.068 & 
   &
   &
    - &
   &
 - &
 - &
 - &
 
 0.081 &
   &
   &
 - &
   &
-  &
 - &
 - \\
 
 &
 Emotion &
 0.153 & 
   &
   &
    - &
   &
 - &
 - &
 - &
 
 0.332 &
   &
   &
 *0.159 &
   &
-  &
 - &
 - \\
 
 &
 Affect &
 0.123 & 
   &
   &
    - &
   &
 - &
 - &
 - &
 
 0.273 &
   &
   &
 - &
   &
-  &
 - &
 - \\
 
\midrule
\multicolumn{18}{l}{\textbf{Note:} \textbf{*} indicates $p<0.05$; \textbf{**} indicates $p<0.01$. $\beta$ values are not shown when not statistically significant.}\\
\bottomrule
\end{tabular}
\end{adjustbox}
\end{table*}

Where:
\begin{itemize}
    \item $M_{LIWC}$ is the measure for a given LIWC category (e.g., Tone)
    \item \(\beta_0\) is the baseline effect present in the core prompt and LLM. 
    \item \(\beta_1, \beta_2, \beta_3\) are coefficients represent the influence of the factors \textit{Attitude}, \textit{Authority}, and \textit{Reasoning} respectively on the linguistic quality. 
    \item \(\beta_4, \beta_5, \beta_6,\beta_7\) represent the interaction effects between the factors. (ie. \(\beta_4 \)  represents the interaction between Attitude and Authority.) 

    \item \(\epsilon\) represents the error term.
\end{itemize}

\section{Results}
Table \ref{tab:LIWC_measures} shows the LIWC categories used per personality dimension. 
A category is correlated with a dimension if it is significantly different on the low and high end of that personality scale, e.g. Tone is significantly correlated with Attitude if it is sufficiently different in optimistic and pessimistic chatbots. 
If a coefficient is positive, it implies that as the factor increases, the linguistic quality (in that particular category) increases.

\section{Discussion}
Our study evaluated the capacity of LLMs to manifest engineered prompted traits. The results offer several key insights:

\subsection{GPT Model Variation} 
GPT-4 appears to be more sensitive to the prompted traits across a broader range of LIWC categories compared to GPT-3.5. 
The wider range of significant categories in GPT-4 might suggest an enhanced capability to capture and reflect certain nuances, but it doesn't automatically imply superior overall performance in expressing the prompted traits. 
The explanation may also lie to some extent with the nature of LIWC dictionary categories. 

Consider the following text generated by a CA powered by GPT-3.5 with the personality traits of optimistic-authoritative-analytical: ``As \textit{someone} who deeply \textit{cares} about the welfare of animals and the environment, I \textit{found} my \textit{purpose} in \textit{helping} organizations like the Wildlife Bridge Foundation.''
Compare this with another CA with the same GPT-3.5 model, but with the personality traits of optimistic-authoritative-affective: ``As \textit{someone} who deeply \textit{cares} about the \textit{well-being} of animals, it breaks my heart to see their habitats \textit{destroyed} by urban expansion.''

In both examples, the italicized terms correspond to not just LIWC categories corresponding to analytical and affective personalities, but also optimistic and pessimistic personalities.
This is because LIWC dictionary categories are not mutually exclusive: the same word can appear in multiple categories. A more thorough evaluation of this approach would also need to incorporate a measure of overlapping words across relevant LIWC categories from the generated text.

\subsection{Conversation Benchmarking}
Our demonstrated approach of iterative simulated conversations has potential to serve as a method for assessing the variability in LLM output within defined prompted traits.
While this approach would need a more rigorous validation of the representation of these personalities, the use of expert-curated psycholinguistic dictionary categories---such as those in LIWC---could provide CA designers with a standard set of linguistic measures to use for specific applications.

\subsection{Trait Variability}
Prompted traits have very different impact on linguistic quality. 
\textit{Attitude} significantly influences linguistic qualities across GPT models. \textit{Authority} has a stronger impact on the GPT-4 model, while \textit{Reasoning} shows the least effect across models.
Identifying which traits have stronger impacts allows developers to better tailor LLM outputs.

Furthermore, we wish to highlight the nuanced relationship between prompt engineering and resultant LLM outputs, especially when a dearth of good methods for evaluating output variability makes it doubly difficult to quantify this relationship.
By demonstrating one method, we hope to bring attention to the remaining scope for refining evaluation and monitoring techniques for quality and consistency of generated content. 


\subsection{Implications for personality-based CAs.}
Given these results, we formulate the following design and development recommendations for effective and consistent personality-based CAs:

\begin{enumerate}[leftmargin=*]
\item \textbf{Prompt Design:} We hypothesize that introducing more synonyms to a given prompt when creating personality archetypes may lend the personality element more weight and create more long-standing effects. 
For example, an optimistic agent may be described as optimistic, positive, and hopeful in a prompt instead of simply optimistic.

\item \textbf{Prompt Programming:} We suggest periodic re-injection of the persona prompt components during the conversation could enforce model adherence to the intended persona, potentially mitigating drift.

\item \textbf{LIWC-Annotated Prompt Programming:} by real-time monitoring present LIWC categories in conversation output, any significant deviation or drift from the desired personality could be detected.
This then can create a feedback loop to nudge the model back to the desired personality traits, ensuring consistent alignment with the initial prompt.

\end{enumerate}

Future research could also explore the role of user input in conversation on steering agent personality over time, the generalizability of our domain-driven personality crafting approaching, and expand the set of LLM models evaluated in this work.

\section{Conclusion}
We demonstrate in this paper the feasibility of generating CA personalities through LLM prompt engineering. We craft personality archetypes from literature and charity solicitation domain knowledge, evaluating the effect of personality prompts in LLM CAs on language style using LIWC computational text analysis. We show that the performance of LLM CAs in this area is model dependant, and personality dimension dependant. Based on our results, we present design and development recommendations to fellow researchers, including recommendations on prompt engineering and future research directions. 


\balance
\bibliographystyle{ACM-Reference-Format}
\bibliography{references}

\end{document}